\documentclass[12pt]{article}
\usepackage{graphicx}

\def\pd{\partial}
\def\a{\alpha}
\def\b{\beta}
\def\dl{\delta}
\def\s{\sigma}

\def\lam{\lambda}
\def\Lam{\Lambda}
\def\hg{{\hat g}}

\def\bfx{{\bf x}}
\def\bfr{{\bf r}}

\def\QG{{\rm QG}}
\def\P{{\rm P}}
\def\D{{\rm D}}
\def\G{{\rm G}}
\def\cl{{\rm cl}}
\def\sq{\sqrt}
\def\e{\hbox{\large \it e}}

\def\fr{\frac}

\def\pp{\prime}

\def\bb{\begin{equation}}
\def\ee{\end{equation}}
\def\bba{\begin{eqnarray}}
\def\eea{\end{eqnarray}}

\begin{document}

\begin{center}
{\bf Quantum Gravity Scenario of Inflation 
based on the CMB Anisotropies}\footnote{
Talks presented by T.Y. at the Sapporo Winter School in Niseko '04, Jan 8-12, 2004, Niseko Annupuri, Hokkaido, Japan;   
K.H. at the 12th International Conference on Supersymmetry and Unification of Fundamental Interactions, June 17-23, 2004, Epochal Tsukuba, Tsukuba, Japan; 
at the 17th International Conference on General Relativity and Gravitation, July 18-23, 2004, 
RDS, Dublin, Ireland.}
\end{center}

\begin{center}
{Ken-ji Hamada$^a$ and Tetsuyuki Yukawa$^b$} 
\end{center}

\begin{center}
{}$^a${\it Institute of Particle and Nuclear Studies, KEK, Tsukuba 305-0801, Japan}
{}$^b${\it Coordination Center for Research and Education, \\ 
The Graduate University for Advanced Studies (Sokendai), Hayama 240-0193, Japan}
\end{center}

\begin{abstract}
Inflationary scenario based on a renormalizable model of conformal gravity is proposed 
and primordial spectrum is derived. The sharp fall off of the angular power 
spectra at low multipoles in the COBE and WMAP observations are explained by a dynamical scale of quantum gravity. At this scale, the universe would make a sharp transition from the quantum spacetime with conformal invariance to the classical spacetime.
\end{abstract}

\vspace{1mm}

\normalsize\baselineskip=15pt

\begin{flushleft}
{\large {\bf 1. Introduction}}
\end{flushleft}

In 2003, the Wilkinson Microwave Anisotropy Probe (WMAP)~\cite{wmap} mission 
released the first year result. One of the significant result is that WMAP reconfirmed  
the sharp fall off of the angular power spectrum at the ultra super-horizon region, 
previously discovered by the Cosmic Background Explorer (COBE)~\cite{cobe}. 
Although the deviation from Harrison-Zel'dovich spectrum~\cite{hz} is usually 
explaind in terms of the cosmic variance, 
this deviation could be explained as an appearance of new dynamical scale, since 
cosmic variance is based on the assumption of ergodicity, although in the super-horizon region  
there is no mixing process required for the statistical treatment. 
Amazingly, if we believe the idea of inflation originally introduced to resolve the flatness 
and horizon problems~\cite{guth}, 
the observed spectrum provides us information about dynamics beyond the Planck scale. 
Our proposal is that at the very early universe there is a sharp transition from the quantum spacetime with the full conformal invariance to the classical Friedmann spacetime, 
and WMAP observes this instance~\cite{hy}.

\begin{flushleft}
{\large {\bf 2. The Model}}
\end{flushleft}

Despite the common belief that there exists no consistent quantum theory of gravity at hand, we propose a model of quantum gravity which explains the sharp fall of the spectrum, and also ignites the inflation naturally without any additional fields.
The model we employ is a renormalizable model 
based on the conformal gravity in 4-dimension defined by the action~\cite{hamada02}
\bb
   I=\int d^4 x \sq{-g} \left\{ 
     -\fr{1}{t^2}C_{\mu\nu\lam\s}^2 -bG_4 + \fr{M_\P^2}{2} R 
     -\Lam_{\rm COS} 
       + \cdots \right\},   
            \label{action}
\ee
where $M_\P =1/\sq{8\pi G}$ is the reduced Planck mass, and $\Lam_{\rm COS}$ is the cosmological constant. $C_{\mu\nu\lam\s}$ is the Weyl tensor, and $G_4$ is the Euler density. 
The dots denote conformally coupled matter fields.

The metric field is decomposed to the conformal mode $\phi$ 
and the traceless mode $h^{\lam}_{~\nu}$ with $tr(h)=0$ as
$g_{\mu\nu} = \e^{2\phi} \hg_{\mu\lam} (\dl^{\lam}_{~\nu} + t  h^{\lam}_{~\nu} +\cdots )$.
The traceless mode will be handled perturbatively in terms of the coupling $t$ on the background metric $\hg_{\mu\nu}$, while the conformal mode is treated {\it non-perturbatively}. 
The beta function of the renormalized coupling $t_r$ has been calculated, which indicates the asymptotic freedom: $\b_t =-\b_0 t_r^3$ with $\b_0 > 0$. 
Thus, the dynamics of the traceless mode introduces the scale parameter $\Lam_\QG$ in the similar manner as the running coupling of gauge theories;
\bb
   \a_\G =\fr{t_r^2(p)}{4\pi}= \fr{1}{4\pi \b_0}\fr{1}{\log (p^2/\Lam_\QG^2)}. 
         \label{alphaG}
\ee  
The appearance of $\Lam_\QG$ will be shown to influence the low multipole components of the angular power spectra.

The order of two mass scales involved in our model that affect the dynamics at the early universe 
are set as $M_\P \gg \Lam_\QG$.
The asymptotic freedom implies at very high energies above the Planck mass 
the Weyl tensor should vanish, and spacetime fluctuations are dominated 
by the conformal mode, and the dynamics is governed by the conformal field theory (CFT$_4$) 
with the full conformal invariance reflecting the background-metric independence.   
There exists no classical spacetime and the universe is filled with quantum fluctuation of the conformal field.

As the space expands the energy gets lowered to $M_\P$, and the Einstein action becomes effective. 
The conformal symmetry is slightly broken, and  
the conformal field fluctuates around a {\it stable} inflating solution $\phi_\cl$ with the 
Hubble constant $H_\D=\sqrt{8\pi^2/b_1}M_{\rm P}$ following the classical equation of motion   
\bb
    -b_1\pd_\eta^4 \phi_\cl   
       +24 \pi^2 M_\P^2 \e^{2\phi_\cl} \left\{ \pd_\eta^2 \phi_\cl 
                                         + (\pd_\eta \phi_\cl)^2 \right\} 
    =0,
         \label{eom}
\ee
depending only on the conformal time $\eta$. 
The first term comes from the Wess-Zumino action for the conformal anomaly induced from 
the diffeomorphism invariant measure. The coefficient is computed as  
$b_1 = ( N_{\rm X} +\fr{11}{2}N_{\rm W} +62 N_{\rm A} )/360 +769/180$, 
where $N_{\rm X}$, $N_{\rm W}$ and $N_{\rm A}$ are the numbers of scalar fields, 
Weyl fermions and gauge fields, respectively.

The inflation will terminate eventually at the energy scale $\Lam_\QG$, where the effective coupling strength to the traceless mode diverges, and the action will depart from the CFT 
action  considerably. At this stage we expect the correlation of field fluctuations becomes 
short range and spacetime becomes classical. 
The universe would make a sharp transition from the inflation era to the Friedmann era. 
The field fluctuation percolates to localized objects, and they eventually decay into the 
ordinary matter driving the universe into the big bang phase. 
The primordial fluctuation we observe in the CMB is expected to be quantum fluctuation of the conformal mode developed right before this transition.

Since the inflation starts at the Planck scale $\tau_\P=1/M_{\rm P}$ and 
ends at $\tau_{\Lam}=1/\Lam_\QG$, where $\tau$ is the proper time defined 
by $d\tau =a(\eta) d\eta$ with $a=\e^{\phi_{\rm cl}}$, the number of e-foldings of inflation is estimated as 
${\cal N}_e = \log a(\tau_{\Lam})/a(\tau_\P) \simeq  M_{\rm P}/\Lam_\QG$. 
Once the value of $\Lam_\QG$ is chosen to fit the observed data, we can check whether the model can give the right number of e-foldings large enough to solve the flatness and horizon problems.

\begin{flushleft}
{\large {\bf 3. Primordial Power Spectrum}}
\end{flushleft}

At large scale, the CMB anisotropies reflect gravitational potential on the last scattering 
surface, known as the Sachs-Wolfe effect~\cite{sw}: $\dl T/T =\Phi({\bf x}_{\rm lss})/3$. 
In order to relate the observed anisotropies to the quantum fluctuation, we consider the process dividing into two steps: the step at the recombination time $\tau_{\rm rec}$, and the step at the dynamical transition $\tau_\Lam$.
At $\tau_{\rm rec}$ the potential fluctuation $\Phi$ is related to the density 
contrast $\dl\rho/\rho$ through the Poisson law.  
At $\tau_\Lam$ we assume the primordial density fluctuation precisely reflects the scalar curvature fluctuation, which is the unique observable with the general covariance:
$\dl R/R \bigr|_{\tau_\Lam^-} ~\sim~~ \dl\rho/\rho \bigr|_{\tau_\Lam^+}$,
where $\tau_\Lam^{-(+)}$ is the time just before(after) $\tau_\Lam$. This density contrast at the large scale with super-horizon separations will be preserved until $\tau_{\rm rec}$.

The two point quantum correlation function of the scalar curvature contrast $\dl R/R$ 
in CFT$_4$ is given by  
$
    \langle\langle  
      \dl R/R (\tau_\Lam, \bfr) \dl R/R (\tau_\Lam, \bfr^\pp) 
        \rangle\rangle 
     \sim \left( H_\D |\bfr -\bfr^\pp | \right)^{-2\Delta_R},
$
where $|\bfr-\bfr^\pp| =a(\tau_\Lam) |\bfx-\bfx^\pp| $ is the physical distance 
on the hypersurface at $\tau_\Lam$, and $\Delta_R $ is the scaling dimension of 
the curvature operator determined from the scale transformation property of the 
operator.  The spectral index is computed as 
$n=2\Delta_R-3=5-8(1-\sqrt{1-2/b_1})/(1-\sqrt{1-4/b_1})$. 
The deviation from the Harrison-Zel'dovich spectrum~\cite{hz}, 
$n-1=2/b_1 + 4/b_1^2 + o\left( 1/b_1^3 \right)$, is the contribution from CFT$_4$.

The two point correlation function involves integration over the momentum, which runs in the region where the coupling with the traceless mode is not negligible. 
Thus, there are corrections from the traceless mode 
and the index has correction from the coupling constant.  
In order to take into account the non-perturbative effect we replace the effective coupling constant by the running coupling $\a_\G$ (\ref{alphaG}).
We then obtain the angular power spectrum for large angles to be
\bb
   C_l 
    = \int^{\infty}_{\lam} \fr{dk}{k}j_l^2(kx_{\rm lss})P (k) \quad 
{\rm with} \quad 
     P(k) = A \left( \fr{k}{m_\lambda} \right)^{n-1+\fr{v}{\log(k^2/\lam^2)}},
\ee
where $A$ is a dimensionless normalization constant and $v$ is a positive constant.
$m_\lam$ and $\lam$ are the comoving Planck and dynamical scales at the time $\tau_{\Lam}$ defined by $m_\lam =a(\tau_\Lam) H_\D$ and $\lam =a(\tau_\Lam) \Lam_\QG$.

We now determine parameters by comparing low multipole components of the angular power spectra 
of the WMAP observation. 
We focus on the sharp damping at $l=2$ and $3$ multipole components as an appearance of the dynamical scale of quantum gravity. 
This determines a value for the comoving dynamical scale to be about $\lam=3/x_{\rm lss}$. 
If we take $x_{\rm lss}=14000$ Mpc, we obtain $\lam=0.0002$ Mpc$^{-1}$. 
The number of e-foldings, ${\cal N}_e \simeq H_\D/\Lam_\QG = m_\lambda/\lam$, can be taken an arbitrarily large value according to the value $m_\lambda$. 
For example, we can choose $m_\lambda=0.02$ Mpc$^{-1}$ so that ${\cal N}_e=100$ which is large enough to solve the flatness problem. 
Since $H_\D$ is order of $10^{19}$ GeV, we have $\Lam_\QG \sim 10^{17}$ GeV.

The spectral index $n$ of the Standard Model ($n=1.41$ from $N_A=12$, $N_W=45$) predicts large blue spectrum. From the observation it looks favorable to make $n$ a little smaller by adding extra matter and gauge fields as the GUT or SUSY models suggest. The integrated Sachs-Wolfe(ISW) effect may shift the index up. 
However, the conclusive statement should be postponed until the overall analysis based on our spectrum is completed.
The CMB spectra, $l(l+1)C_l$, are calculated up to $l = 40$ for $n=1.1$, $1.2$, $1.3$ as shown in 
figure in company with the WMAP data.   
The normalization constant $A$ is chosen appropriately so that $l=6$ multipole components coincide with the observed value, and $v$ is taken to be $0.001$.
\begin{figure}
\begin{center}
\begin{picture}(300,200)(100,0)
\put(50,0){\includegraphics[width=12cm,clip]{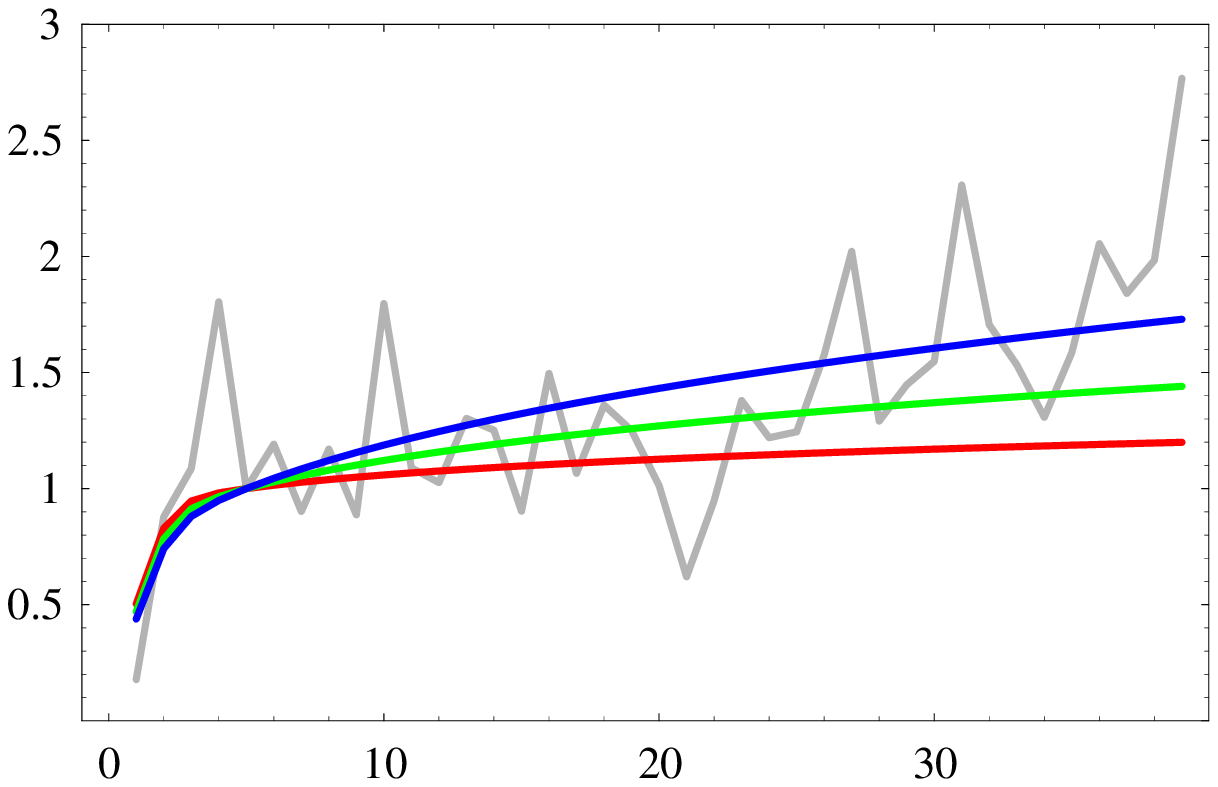}}
\put(45,100){\rotatebox{90}{$l(l+1)C_l$}}
\put(240,-10){$l$}
\put(385,90){$n=1.1$}
\put(385,105){$n=1.2$}
\put(385,125){$n=1.3$}
\put(385,160){{\rm WMAP data}}
\end{picture}
\end{center}
\end{figure}

In summary, we have shown an inflationary scenario induced by quantum gravity and derived the CMB angular power spectrum at large angles.  
The sharp damping at low multipole components is interpreted as the reflection of 
the dynamical scale of quantum gravity. 
Since the primordial spectrum is produced by CFT$_4$, we expect that the tensor-to-scalar ratio is negligible and multi point correlation functions of non-Gaussian type exist.    

\bibliographystyle{plain}

\end{document}